\documentstyle[prl,aps,epsf,floats,twocolumn]{revtex}
\ifpreprintsty \else    \fi
\begin{document}
\draft
\ifpreprintsty \else \wideabs{ \fi
\title{Direct calculation of the hard-sphere crystal/melt interfacial 
free energy}
\author{Ruslan L. Davidchack and Brian B. Laird}
\address{Department of Chemistry and Kansas Institute for Theoretical
and Computational Sciences, University of Kansas, Lawrence, Kansas 
66045}
\date{\today}
\maketitle
\begin{abstract}
We present a direct calculation by molecular-dynamics computer 
simulation of the crystal/melt interfacial free energy, $\gamma$, for a
system of hard spheres of diameter $\sigma$.
The calculation is performed by thermodynamic integration along a 
reversible path defined by cleaving, using specially constructed 
movable hard-sphere walls, separate bulk crystal and fluid systems,
which are then merged to form an interface.  We find the interfacial
free energy to be slightly anisotropic with $\gamma$ = 0.62$\pm 0.01$, 
0.64$\pm 0.01$ and 0.58$\pm 0.01 k_BT/\sigma^2$ for the (100), (110) 
and (111) fcc crystal/fluid interfaces, respectively.  These values 
are consistent with earlier density functional calculations and
recent experiments measuring the crystal nucleation rates from 
colloidal fluids of polystyrene spheres that have been interpreted 
\symbol{91}Marr and Gast, Langmuir {\bf 10}, 1348 (1994)\symbol{93}
to give an estimate 
of $\gamma$ for the hard-sphere system of 
$0.55 \pm 0.02 k_BT/\sigma^2$, slightly lower than 
the directly determined value reported here.
\end{abstract}
\ifpreprintsty \else } \fi

A detailed microscopic description of the interface
between a crystal and its melt is necessary for a full understanding of
such important phenomena as homogeneous nucleation and
crystal growth\cite{Woodruff73,Howe97,Adamson97}. Computer 
simulation studies of model materials
have had some success in elucidating the phenomenology of such 
systems\cite{Laird98}, the importance of such work being enhanced by
the near absence of reliable experimental studies.  These efforts, 
however, have been primarily focused on structural and dynamical 
properties, since the central thermodynamic property, the
crystal/melt interfacial free energy, is difficult
to measure by simulation {\it or} experiment. In this work we
report the results of a {\it direct} calculation via molecular-dynamics
(MD) simulation of the crystal/melt surface free energy of the 
hard-sphere system, one of the most important reference models for 
simple materials.

The crystal/melt surface free energy, $\gamma$, is defined as the 
(reversible) work required to form a unit area of interface between
a crystal and its coexisting melt. 
Experimentally, $\gamma$ can be measured either indirectly from 
measurements of crystal nucleation rates interpreted through classical 
nucleation theory, or directly by contact angle 
measurements \cite{Woodruff73}.  Using the former method, 
Turnbull \cite{Turnbull50} estimated $\gamma$ for a number of 
materials and found a strong empirical correlation between the values 
obtained and the latent heat of fusion for each material given by the 
relation $\gamma \approx C_T \Delta_f H \rho^{2/3}$, where $\rho$ is 
the number density of the crystal and with $C_T$ (the Turnbull 
coefficient) taking on the value 0.45 for most metals and 0.32 for 
other mostly nonmetallic materials. 
For the hard-sphere system considered in this work, recent 
experiments\cite{Dhont92} of the crystallization kinetics of a 
colloidal suspension of uncharged polystyrene spheres, 
which closely mimic hard spheres, have been interpreted within a 
classical nucleation model to yield an estimate for $\gamma$ of the 
hard-sphere system of 0.55$\pm 0.02$$k_BT/\sigma^2$ \cite{Marr94}. 
This value is in agreement \cite{Marr95} with that predicted using
the empirical relationship above assuming a $C_T$ of 0.45 and values of
$\Delta_f H$ and coexistence densities for hard spheres as determined
by MD simulation \cite{Hoover68}. Unfortunately, the accuracy of 
the values of $\gamma$ obtained from nucleation rates is severely 
limited by the approximations inherent in classical nucleation theory.
More accurate values can be obtained directly using contact angles, 
but such experiments are difficult and only a few materials have been 
studied to date.  One notable example is a series of grain boundary 
contact angle experiments on bismuth \cite{Glicksman69} that determined 
$\gamma$ to be relatively independent of crystal orientation 
at 61.3$\times 10^{-3}$ J/m$^2$, which is about 10\% higher than 
Turnbull's estimate from nucleation rates of 
54.4$\times 10^{-3}$ J/m$^2$. 

In recent years, the primary theoretical approach to studying the 
structure and thermodynamics of the crystal/melt interface has been 
density-functional theory (DFT)
\cite{McMullen88,Oxtoby88,Curtin89,Marr93,Kyrlidis95,Ohnesorge95}. 
For these studies, the hard-sphere system has been the benchmark 
calculation, due to the simplicity of the interaction and 
the availability of accurate, analytical formulas for the properties 
of the fluid. However, as discussed by Marr \cite{Marr95} the value 
of $\gamma$ obtained is  highly dependent on the approximations used 
to construct the DFT and the reported values range from
0.25 to 4.00$k_BT/\sigma^2$.  The DFT studies also disagree 
dramatically in the degree of orientation dependence of the 
interfacial free energy.  Unfortunately, in the absence of simulation 
results it has been difficult to assess the validity of the individual 
approaches, although only the DFT approach
of Curtin\cite{Curtin89} ($\gamma = 0.62 k_BT/\sigma^2$)  
and the related one of Marr and Gast\cite{Marr94} 
($\gamma = 0.60 k_BT/\sigma^2$) come close to the nucleation result 
of $0.55 \pm 0.02 k_BT/\sigma^2$.

To date, the only reliable calculation of the 
crystal/melt interfacial free energy via simulation is 
that of a system of particles interacting via a truncated Lennard-Jones 
potential by Broughton and Gilmer \cite{Broughton86vi}. In that work, a 
series of continuous, external ``cleaving'' potentials are used to 
separate (cleave) separate samples of bulk liquid and fcc crystal, 
prepared at the calculated coexistence temperature and densities. 
The solid and liquid slabs thus produced are then placed next to one 
another and the cleaving potentials removed to merge them into a 
coexisting interface. The reversible work to perform these steps can 
be calculated by thermodynamic integration, giving a direct calculation
of $\gamma$ for this system.  The values of $\gamma$ at the triple 
point were found to be statistically isotropic
with $\gamma \sigma^2/\epsilon = 0.35\pm 0.02, 0.34\pm 0.02$ and 
$0.36\pm 0.02$ for the (111), (100) and (110) interfaces, respectively.

For the hard-sphere system, the Broughton-Gilmer procedure must be
modified since the algorithm for MD simulation of discontinuous 
hard-core potentials is conceptually very different from the algorithm 
for continuous potentials.  The latter are performed by integrating 
the system of ordinary differential equations, while in the former 
the dynamical algorithm proceeds on a collision by collision basis.
Therefore, incorporating continuous cleaving potentials into the 
collisional algorithm would result in lost efficiency and substantial 
modification of the structure of the algorithm.

\begin{figure}
  \ifpreprintsty \vspace*{0pt} \hspace*{0pt}  \epsfxsize=360pt
  \else          \vspace*{0pt} \hspace*{0pt}  \epsfxsize=235pt  \fi
  \epsfbox{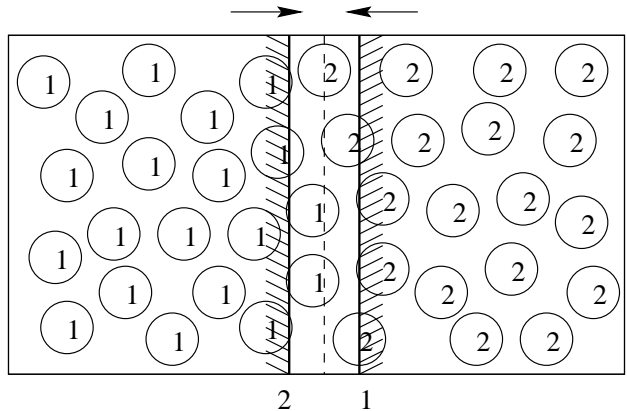}
  \ifpreprintsty \par\vspace*{0pt}  \else \par\vspace*{10pt}  \fi
\caption{Diagram illustrating the cleaving of the bulk hard-sphere
system by two moving walls.  Spheres are assigned types 1 and 2 based 
on their position with respect to the cleaving plane (dashed line).
Two walls of types 1 and 2, which interact only with spheres of similar
type, are placed on the opposite sides of the cleaving plane, so that
initially there are no collisions between walls and spheres (as shown
on the diagram).  The system is then cleaved by moving the walls in
directions indicated by the arrows.}
\label{fig:diagram} \end{figure}

In this Letter we introduce an approach, which uses only hard-sphere
interactions in order to cleave the bulk hard-sphere systems.
This allows us to apply the Broughton-Gilmer cleaving procedure to the
hard-sphere system with only minor changes to the algorithm structure.
The idea of our approach is illustrated in Fig.~\ref{fig:diagram}.
To cleave the bulk system at a {\em cleaving plane} (shown by the
dashed line in Fig.~\ref{fig:diagram}), the spheres are assigned 
types 1 or 2 based on their position relative to the plane.
Next, two walls, which are also assigned types 1 and 2, are placed
on the opposite sides of the cleaving plane.  The two types are
introduced in order to specify interaction between the spheres and 
the walls; namely, {\it  the walls interact only with the spheres of
similar type}.  Therefore, when the walls are placed as shown in
Fig.~\ref{fig:diagram}, and the distance from the walls to the
cleaving plane is larger than the sphere radius, the walls do
not interact with the spheres.
It is important that, during a simulation run, a sphere 
changes its type whenever it crosses the cleaving plane.
Because of the periodic boundary conditions in the $z$ direction, 
another plane must be defined sufficiently far away from the cleaving
plane, where the spheres also change type. 

The cleaving of the system is achieved by slowly moving the walls
towards each other (as shown by the arrows in Fig.~\ref{fig:diagram}),
starting from the initial position $z_i$, where the walls do not
interact with the system, till $z_f$, where the spheres of different 
types no longer collide with each other at the cleaving plane.  
During the process, the average pressure on the walls, $P(z)$, is 
measured as a function of wall position.  The work per unit area 
required to perform the cleaving is given by the integral
\begin{equation}
  w = \int^{z_f}_{z_i} P(z)\;dz\:.
  \label{eq:work} \end{equation}
Thus the crystal-fluid interfacial free energy, $\gamma$, can be
measured in the {\em reversible} process involving the following
four steps: (1) Cleave the bulk crystal by inserting two walls at
the cleaving plane and moving them from $z_i$ to $z_f$; (2) Cleave the
bulk fluid in a similar way; (3) Juxtapose the cleaved crystal and
fluid systems by changing the periodic boundary conditions while
retaining the crystal and fluid systems restricted by the respective
cleaving walls; (4) Slowly move the walls back to their initial
positions with respect to the cleaving planes.
This series of  steps is the same as that used by Broughton and 
Gilmer\cite{Broughton86vi},
except that, in our case, no work is done on the system in step 3.
The interfacial free energy is given by the sum 
$\gamma = w_1 + w_2 + w_4\;$,
where $w_4$ is negative and consists of the work done by the walls
on the crystal and fluid parts of the system during the process of
removing the cleaving walls.

The structure of the cleaving walls is crucial to the success of the
procedure.  The main requirement is that the insertion of the walls
perturbs the systems as little as possible.  Our approach is to use
walls made of layers of ideal crystal oriented in correspondence with
the orientation of the crystal system.  For the (100) and (111) 
orientations it is sufficient to use one layer, while for the (110) 
orientation we use two layers.  Such a choice of the wall structure
ensures minimal perturbation of the cleaved crystal, while the 
cleaved fluid is expected to form properly oriented interfacial layers.
On a technical side, the implementation of such a wall structure
is quite simple, since we can treat collisions with the walls in the
same manner we treat collisions between all the spheres in the system,
except that the spheres forming the walls are assigned infinite mass.

\begin{figure}
  \ifpreprintsty \vspace*{0pt} \hspace*{0pt}  \epsfxsize=360pt
  \else          \vspace*{0pt} \hspace*{0pt}  \epsfxsize=235pt  \fi
  \epsfbox{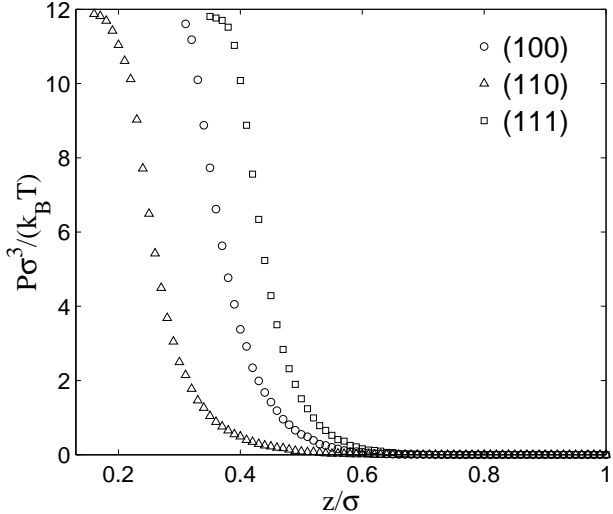}
  \ifpreprintsty \par\vspace*{0pt}  \else \par\vspace*{10pt}  \fi
\caption{Step 1: Cleaving crystal system.  Pressure on the cleaving 
walls as a function of wall position for the three orientations
of the interface.  The error bars are smaller than the
size of the symbols.}
\label{fig:crystal} \end{figure}

The position of the walls, $z$, is measured by the distance of the
centers of the spheres forming the walls to the cleaving plane.
Obviously, the walls do not interact with the system when $z > \sigma$.
Therefore, we set the initial position of the walls at $z_i = \sigma$.
The pressure $P(z)$ is obtained by
moving the walls from $z_i$ to $z_f$ in steps of 0.01$\sigma$.
In order to move the walls to a new position, we assign a small 
velocity (about 0.1\% of the average particle velocity) to the spheres 
forming the walls, and run the simulation until the walls reach the 
new position.  At that moment the wall velocity is set to zero, and the
velocities of the particles are rescaled in order to restore the
initial value of the total kinetic energy of the system.
Before measuring the pressure, we allow the system to relax in an
equilibration run.
In order to verify the reversibility of the cleaving process, we have
also simulated the reverse process and measured the pressure while
moving the walls from $z_f$ back to $z_i$.  The details of the
simulation process and obtained results follow.

{\em Step 1: Cleaving the crystal} ~For each of the three orientations, 
we start with a crystal at a density $\rho_c = 1.037\sigma^{-3}$, 
which corresponds to the crystal-fluid coexistence pressure of 
11.55$k_B T \sigma^{-3}$\cite{Davidchack98}.
In order to minimize the size effects and the amount of stress
in the crystal introduced by the cleaving walls,  
we use large systems of about 8000 spheres and approximate 
dimensions of $14\sigma \times 14\sigma \times 40\sigma$.
(To perform the simulations efficiently for such large systems, 
we use the algorithm of Rappaport \cite{Rappaport95}.)
The cleaving plane is placed in the middle between two crystal layers. 
The dependence of the pressure on the wall position is shown in
Fig.~\ref{fig:crystal}.  The walls do not interact with the crystal
until they move sufficiently close to the crystal layers 
(about 0.7$\sigma$ for all orientations).  Then the pressure on the
walls quickly rises to slightly above the bulk crystal pressure.
The steepness of the rise is directly related to the compactness
of the layers for each orientation.
The final positions, $z_f$, where the spheres of different types
no longer collide across the cleaving plane, were determined to be 
0.31$\sigma$, 0.16$\sigma$, and 0.35$\sigma$ for the (100), (110), 
and (111) system orientations, respectively.  No hysteresis was
observed in the reverse process.

\begin{figure}
  \ifpreprintsty \vspace*{0pt} \hspace*{0pt}  \epsfxsize=360pt
  \else          \vspace*{0pt} \hspace*{0pt}  \epsfxsize=235pt  \fi
  \epsfbox{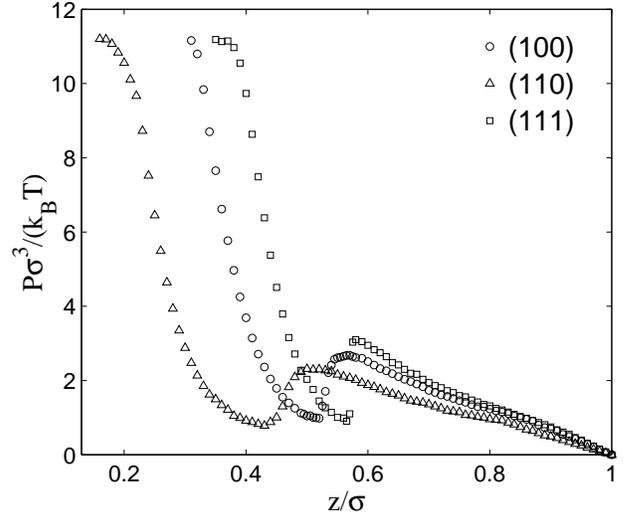}
  \ifpreprintsty \par\vspace*{0pt}  \else \par\vspace*{10pt}  \fi
\caption{Step 2: Cleaving fluid system.  Pressure on the cleaving 
walls as a function of wall position for the three orientations
of the interface.  The error bars are of the order of the
size of the symbols.}
\label{fig:fluid} \end{figure}

{\em Step 2: Cleaving the fluid} ~The fluid systems consisting of about
7400 particles are prepared at the coexistence density 
$\rho_f = 0.939\sigma^{-3}$ using box dimensions nearly identical to 
the crystal systems.  Unlike in step 1, the cleaving walls begin to 
interact with the fluid system as long as $z < \sigma$.  As can be 
seen in Fig.~\ref{fig:fluid}, the pressure on the walls
increases approximately linearly until the 
fluid near the cleaving walls begins to develop significant crystal-like
ordering commensurate with the wall structure.  At that point
the pressure in the bulk fluid decreases to about 
11.2$k_B T \sigma^{-3}$, after which the dependence
of pressure on the wall position follows essentially the same curve 
as during the cleaving of the crystal system, which leads to the same 
values of $z_f$ as in step 1.

Simulation of the reverse process shows that the ordering of the fluid
against the walls is the source of some hysteresis.  However, we have 
found that the magnitude of the hysteresis can be always reduced to 
within the statistical error by increasing the duration of the 
equilibration run.  In other words, at every position of the cleaving 
walls, the pressure on the walls eventually converges to the same value
(within the statistical error bounds) in both forward and reverse 
processes.


{\em Step 3: Changing boundary conditions} ~The combined system has 
two cleaving planes and four walls.  The crystal part of the system 
and the two walls restraining it are assigned type 1, while the fluid 
part and the other two walls are assigned type 2.  

\begin{figure}
  \ifpreprintsty \vspace*{0pt} \hspace*{0pt}  \epsfxsize=360pt
  \else          \vspace*{0pt} \hspace*{0pt}  \epsfxsize=235pt  \fi
  \epsfbox{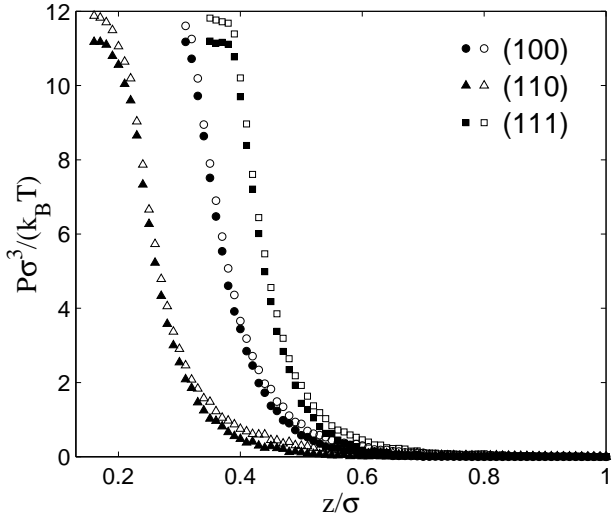}
  \ifpreprintsty \par\vspace*{0pt}  \else \par\vspace*{10pt}  \fi
\caption{Step 4:  Removing the cleaving walls.  Pressure on the 
cleaving walls as a function of wall position for the three orientations
of the interface.  Filled and open symbols indicate pressure on the
walls restraining fluid and crystal parts of the system, respectively.
The error bars are of the order of the size of the symbols.}
\label{fig:interface} \end{figure}

{\em Step 4: Removing the cleaving walls} ~As can be seen in 
Fig.~\ref{fig:interface}, the pressure on the walls restraining 
crystal and fluid parts of the system is essentially the same as in 
steps 1 and 2, respectively, except that the fluid part retains
its structure in the interfacial region. 
Thus the main contribution to the interfacial free energy comes from
the pressure of the fluid on the cleaving walls before significant
crystal-like ordering at the wall develops.

The work done during each step and the resulting interfacial free
energy for each of the three orientations is given in 
Table~\ref{tab:works}. The average values of $\gamma$ 
for the three orientations is about 
0.61$k_BT/\sigma^2$, which corresponds to a Turnbull coefficient of
0.51.  This average value is about 10\% higher than the value 
determined from nucleation rates on colloidal crystals, consistent
with the differences found in other materials, such as bismuth 
(discussed above).  Note that the hard-sphere $\gamma$ values are 
slightly anisotropic and increase in the order of (111), (100) and 
(110).  That (111) has the lowest interfacial free energy is perhaps 
not surprising, since the (111) crystal face resembles most the 
structure adopted by the fluid against a structureless 
wall \cite{Davidchack98}. 

It is generally accepted that the structure and thermodynamics of
dense simple fluids is dominated by the repulsive part of the 
potential, which can often be well approximated as a hard sphere, and
the effect of the attractive  part of the potential can be viewed as
a small perturbation. If one considers the truncated Lennard-Jones
system studied by Broughton and Gilmer and calculates an effective
hard-sphere diameter at the triple point ($T = 0.617\epsilon/k_B$),
using the Barker-Henderson approach\cite{Barker67}, one obtains a value
of ($0.39\epsilon/\sigma^2$) simply by rescaling the hard-sphere result
calculated here.  Thus, the attractive part of this potential accounts
for only about 10\% of the total $\gamma$, which is consistent with a
similar observation by Curtin on the basis of DFT 
calculations \cite{Curtin89}. 

To summarize, we have determined the crystal/melt interfacial free 
energy, $\gamma$ for the hard-sphere system directly from simulation 
using a method that is similar to that Broughton and 
Gilmer \cite{Broughton86vi} used for the truncated Lennard-Jones
system except that we have replaced their external cleaving 
potentials with  specially constructed cleaving walls.  Although the 
method of cleaving walls is especially advantageous for the hard-sphere
system, it could also be easily applied in modified form to continuous 
potentials.  The hard-sphere $\gamma$ obtained is only slightly 
dependent upon orientation and averages about 0.61$k_BT/\sigma$, 
consistent with some previous theoretical predictions from 
density-functional theory.  This value is also about 10\% higher than 
that determined from nucleation experiments on colloidal suspensions 
of polystyrene spheres, giving a rare comparison between direct 
evaluations of $\gamma$ and less accurate indirect determinations via
nucleation theory. 

The authors gratefully acknowledge support from NSF under 
Grant CHE9970903, as well as the Kansas Center for Advanced Scientific 
Computing for computational support.

\begin{table}
\caption{Values of $w_n$ for the steps of the thermodynamic integration
process together with their sum $\gamma$.  The statistical error
estimates represent $2\sigma$ error bounds.}
\begin{tabular}{crrr}
    & $(100)\qquad$ & $(110)\qquad$ & $(111)\qquad$ \\ \hline
$w_1$ & $0.850\pm 0.001$ & $1.287\pm 0.001$ & $1.125\pm 0.001$ \\
$w_2$ & $1.561\pm 0.008$ & $1.989\pm 0.007$ & $1.768\pm 0.008$ \\
$w_4$ & $-1.789\pm 0.005$ & $-2.639\pm 0.006$ & $-2.311\pm 0.005$ \\
$\gamma$ & $0.62\pm 0.01~\;$ & $0.64\pm 0.01~\;$ & $0.58\pm 0.01~\;$
\end{tabular}
\label{tab:works}  \end{table}


\begin{thebibliography}{99}
\bibitem{Woodruff73}
D.P. Woodruff, {\em The Solid-Liquid Interface}, (Cambridge
University Press, London, 1973).

\bibitem{Howe97}
J.M. Howe, {\em Interfaces in Materials}, (John Wiley \& Sons, New
York, 1997).

\bibitem{Adamson97}
A.W. Adamson and A.P. Gast, {\em Physical Chemistry of Surfaces}, 
(Wiley, New York, 1997).

\bibitem{Laird98}
B.~B. Laird, ``Interfaces: Liquid-Solid'' in {\it The Encyclopedia of
Computational Chemistry}, P.v.R Schleyer, N.L. Allinger, T. Clark, 
P. Kollman and H.F. Schaefer, eds. (J. Wiley and Sons, New York, 1998). 

\bibitem{Turnbull50}
D. Turnbull, J. Appl. Phys. {\bf 21}, 1022 (1950).

\bibitem{Dhont92}
J.K.G. Dhont, C. Smits, H.N.W. Lekkerkerker, J. Colloid Interface Sci.
{\bf 152}, 386 (1992). 

\bibitem{Marr94} 
D.W. Marr and A.P. Gast, Langmuir {\bf 10}, 1348 (1994).

\bibitem{Marr95} 
D.W. Marr, J. Chem. Phys. {\bf 102}, 8283 (1995).

\bibitem{Hoover68}
W.G. Hoover and F.H. Ree, J. Chem. Phys. {\bf 49}, 3609 (1968). 

\bibitem{Glicksman69} 
M.E. Glicksman and C. Vold, Acta Met. {\bf 17}, 1 (1969).

\bibitem{McMullen88}
W.E. McMullen and D.W. Oxtoby, J. Chem. Phys. {\bf 88}, 1967 (1988).

\bibitem{Oxtoby88} 
D.W. Oxtoby and W.E. McMullen, Phys. Chem. Liq. {\bf 18}, 97 (1988).

\bibitem{Curtin89}
W.A. Curtin, Phys. Rev. B {\bf 39} 6775 (1989). 

\bibitem{Marr93}
D.W. Marr and A.P. Gast, Phys. Rev. E, {\bf 47}, 1212 (1993). 

\bibitem{Kyrlidis95}
A. Kyrlidis and R.A. Brown, Phys. Rev. E, {\bf 51} 5832 (1995). 

\bibitem{Ohnesorge95}
R. Ohnesorge and H. L\"owen and H. Wagner, Phys. Rev. E {\bf 50} 4801 
(1995). 

\bibitem{Broughton86vi}
J.~Q. Broughton and G.~H. Gilmer, J. Chem. Phys. {\bf 84}, 5759 (1986).

\bibitem{Davidchack98}
R.~L. Davidchack and B.~B. Laird, J. Chem. Phys. {\bf 108}, 9452 (1998).

\bibitem{Rappaport95}
D.C. Rappaport, {\em The Art of Molecular Dynamics Simulation}, 
(Cambridge University Press, New York, 1995). 

\bibitem{Barker67}
J.A. Barker and D. Henderson, J. Chem. Phys. {\bf 47}, 4714 (1967).

\end{thebibliography}
\end{document}